\documentclass[pra,onecolumn]{revtex4}
\usepackage{amssymb}
\usepackage{amsmath}
\usepackage{graphicx}
\usepackage{subfigure}
\usepackage{epstopdf}
\usepackage{color}
\usepackage{verbatim}
\usepackage{titletoc}
\usepackage[hidelinks]{hyperref}

\begin{document}

\title{Suppression of the quasi-two-dimensional quantum collapse in the
attraction field by the Lee-Huang-Yang effect}
\author{Elad Shamriz$^{1}$, Zhaoping Chen$^{1}$}
\email{viskolczp@gmail.com}
\author{ Boris A. Malomed$^{1,2}$}
\address{$^{1}$Department of Physical Electronics, School of Electrical
Engineering, Faculty of Engineering, and
Center for Light-Matter Interaction, Tel Aviv University, P.O.B.
39040, Tel Aviv, Israel\\
$^2$Instituto de Alta Investigaci\'{o}n, Universidad de Tarapac\'{a}, Casilla 7D, Arica,
Chile}

\begin{abstract}
Quantum collapse in three and two dimensions (3D and 2D) is induced by
attractive potential $\sim -r^{-2}$. It was demonstrated that the mean-field
(MF) cubic self-repulsion in the 3D\ bosonic gas suppresses the collapse and
creates the missing ground state (GS). However, the cubic nonlinearity is
not strong enough to suppress the 2D collapse. We demonstrate that the
Lee-Hung-Yang (LHY) quartic term, induced by quantum fluctuations around the
MF state, is sufficient for the stabilization of the 2D gas against the
collapse. By means of numerical solution of the Gross-Pitaevskii equation
including the LHY term, as well as with the help of analytical methods, such
as expansions of the wave function at small and large distances from the
center and the Thomas-Fermi approximation, we construct stable GS, with a
singular density, $\sim r^{-4/3}$ but convergent integral norm.
Counter-intuitively, the stable GS exists even if the external potential is
repulsive, with the strength falling below a certain critical value. An
explanation to this finding is given. Along with the GS, singular vortex
states are produced too, and their stability boundary is found analytically.
Unstable vortices spontaneously transform into the stable GS, expelling the
vorticity to periphery.
\end{abstract}

\maketitle

\section{Introduction and the model}

The quantum collapse, alias \textquotedblleft fall onto the center" \cite{LL}%
, is a well-known phenomenon in quantum mechanics: nonexistence of the
ground state (GS) in three- and two-dimensional (3D and 2D) Schr\"{o}dinger
equations with attractive potential
\begin{equation}
U(r)=-\left( U_{0}/2\right) r^{-2},  \label{U}
\end{equation}%
where $r$ is the radial coordinate, and positive $U_{0}$ is the strength of
the pull to the center. Note that, under the action of potential (\ref{U}),
classical particle of mass $M$ performs motion with angular frequency%
\begin{equation}
\omega =\sqrt{U_{0}/M}r^{-2}  \label{omega}
\end{equation}%
along a circular orbit with arbitrary radius $r$.

In 3D, the collapse occurs when $U_{0}$ exceeds a finite critical value [$%
\left( U_{0}\right) _{\mathrm{coll}}=1/4$ in the notation adopted below],
while in two dimensions $\left( U_{0}\right) _{\mathrm{coll}}=0$, i.e., the
2D collapse happens at any $U_{0}>0$. In both 3D and 2D cases, the potential
represents attraction of a particle (small molecule), carrying a permanent
electric dipole moment, to a central charge, assuming that the local
orientation of the dipole is fixed by the minimization of its energy in the
external field \cite{HS1}. In addition to that, in the 2D case the same
potential (\ref{U}) may be realized as the attraction of a magnetically
polarizable atom to a thread carrying electric current (e.g., an electron
beam) transversely to the system's plane, or the attraction of an
electrically polarizable atom to a uniformly charged transverse thread.
Other 2D settings in Bose-Einstein condensates (BECs) under the action of
similar fields were considered in Refs. \cite{Austria1} and \cite{Austria2}.

A fundamental issue is regularization of the setting, aiming to create a
missing GS. One solution was proposed in Refs. \cite{QTF1}-\cite{QTF3},
which replaced the original quantum-mechanical problem by one based on a
linear quantum field theory. A solution of the latter model produces a GS,
but it does not answer a natural question, what the size of the GS is for
given parameters of the setting, such as $U_{0}$ and mass of the quantum
particle, $M$. In fact, the solution defines the GS size as an arbitrary
spatial scale, which varies as a parameter of the respective quantum-field
renormalization group.

Another solution was proposed in Ref. \cite{HS1}, replacing the 3D linear
Schr\"{o}dinger equation by the Gross-Pitaevskii (GP) equation \cite{GP} for
a gas of the dipole particles pulled to the center by Coulomb potential (\ref%
{U}), and stabilized by repulsive inter-particle interactions. Further, in
the mean-field (MF) approximation which, in particular, assumes the
interaction of the dipole moment of each particle with the electrostatic
field created, as per the Poisson equation, by all other dipoles, the
dipole-dipole interactions between the particles amount to an extra local
cubic term, added to the contact repulsive interaction \cite{HS1}. As a
results, it was found that, in the framework of the MF\ approximation, the
three-dimensional GP equation creates the missing GS for arbitrarily large $%
U_{0}$. The size of this GS is fully determined by parameters of the
physical system ($U_{0}$, $M$, the scattering length of the inter-particle
interactions, and the number of particles, $N$), being $\sim $ a few $%
\mathrm{\mu }$m for typical values of the physical parameters. Furthermore,
beyond the bounds of the MF consideration, it was demonstrated that, in
terms of the many-body quantum theory, treated by means of the variational
Monte-Carlo method, the GS, strictly speaking, does not exist in the same
setting (the collapse is still possible), but the interplay of the pull to
the center and inter-particle repulsion gives rise to a metastable state,
which, for a sufficiently large $N$, is virtually tantamount to the GS,
being separated from the collapsing state by a tall potential barrier \cite%
{Gregory}. Subsequently, the mean-field GS was also found in the 3D gas of
dipole molecules embedded in a strong uniform electric field, which reduces
the underlying symmetry from spherical to cylindrical \cite{HS2}, and in the
two-component 3D\ gas \cite{HS3}, see also a brief review in Ref. \cite%
{cond-matter}.

The situation is more problematic in 2D, as the usual self-defocusing cubic
nonlinearity, which represents the two-body inter-atomic repulsion in the MF
approximation \cite{GP}, is not strong enough to suppress the collapse and
create the GS. The problem is that the MF wave function, $\Psi (r)$,
produced by the respective GP equation, gives rise to the density, $%
\left\vert \Psi (r)\right\vert ^{2}$, diverging $\sim r^{-2}$ at $%
r\rightarrow 0$, in 3D and 2D alike, as this form of the solution is
supported by the balance between the kinetic-energy, pulling-potential, and
cubic terms in the GP equation. Then, in terms of the integral norm,%
\begin{equation}
N=\left( 2\pi \right) ^{D-1}\int_{0}^{\infty }\left\vert \Psi (r)\right\vert
^{2}r^{D-1}dr,  \label{N}
\end{equation}%
where $D=3$ or $2$ is the dimension, the density singularity $\sim r^{-2}$
is integrable in 3D, while it gives rise to a logarithmic divergence in 2D:%
\begin{equation}
N\sim \ln \left( r_{\mathrm{cutoff}}^{-1}\right) ,  \label{ln()}
\end{equation}%
where $r_{\mathrm{cutoff}}$ is the cutoff (smallest) radius. The analysis of
the GP equation readily demonstrates that a self-repulsive nonlinear term
stronger than cubic, i.e., $|\Psi |^{\alpha -1}\Psi $ with $\alpha >3$,
gives rise to the asymptotic form of the density
\begin{equation}
|\Psi (r)|^{2}\sim r^{-4/(\alpha -1)}.  \label{alpha}
\end{equation}%
Thus, any value $\alpha >3$ provides convergence of the 2D integral norm,
given by Eq. (\ref{N}) with $D=2$. In 3D, Eqs. (\ref{alpha}) and (\ref{N})
demonstrate that the critical value of the repulsive-nonlinearity power,
which also entails the logarithmic divergence [cf. Eq. (\ref{ln()})], is $%
\alpha =7/3$ (it is relevant to mention that $\alpha =7/3$ corresponds to
the effective repulsion in the density-functional model of the Fermi gas
\cite{Fermi2}-\cite{Fermi3}).

Thus, a solution for the regularization of the 2D setting may be offered by
the quintic defocusing nonlinearity \cite{HS1}, with $\alpha =5$. The
quintic term accounts for three-body repulsive interactions in the bosonic
gas \cite{Abdullaev1,Abdullaev2}, although an essential difficulty in the
physical realization of the latter feature is the fact that three-particle
collisions give rise to effective losses, by kicking out particles from BEC
to the thermal halo \cite{loss1}-\cite{loss3}.

Recently, much interest was drawn to the formation of quasi-2D and 3D
self-trapped states in BEC in the form of \textquotedblleft quantum
droplets", filled by an effectively incompressible binary condensate, which
is considered as an ultradilute quantum fluid. This possibility was
predicted in the framework of the 3D \cite{Petrov} and 2D \cite%
{Petrov-Astra,Zin,Santos} GP equations which include the Lee-Huang-Yang
(LHY) corrections to the MF approximation, that represent effects of quantum
fluctuations around the MF states \cite{LHY}. The binary structure of the
underlying condensate is essential because the nearly complete cancellation
between the inter-component MF attraction and intra-component repulsion
(which may be adjusted by means of the Feshbach-resonance technique \cite%
{Feshbach})\ makes it possible to create stable droplets through the balance
of the LHY-induced higher-order (quartic) self-repulsion and the relatively
weak residual MF attraction, accounted for by the cubic term. The so
predicted quantum droplets were created with quasi-2D (oblate) \cite%
{Leticia1,Leticia2} and fully 3D (isotropic) \cite{Inguscio1,Inguscio2}
shapes in a mixture of two different spin states of $^{39}$K atoms, as well
as in a heteronuclear mixture of $^{41}$K and $^{87}$Rb atoms \cite{hetero}.
Further, it was theoretically predicted that 2D \cite%
{Raymond1,Raymond2,lattice} and 3D \cite{Barcelona} droplets with embedded
vorticity have their well-defined stability regions too. The LHY effect also
helps to create stable 3D droplets in single-component BECs with long-range
interactions between atoms carrying magnetic dipole moments, as was
demonstrated experimentally and studied in detail theoretically \cite{Pfau1}-%
\cite{Pfau5}, although dipolar-condensate droplets carrying embedded
vorticity are unstable \cite{Macri}.

The objective of the present work is to make use of the LHY effect for the
stabilization of the GS in the quasi-2D bosonic gas pulled to the center by
potential (\ref{U}). This possibility is essential because, as said above,
the alternative, in the form of the quintic self-repulsion, is problematic
in the real BEC setting. The underlying full (3D) GP equation, including the
LHY-induced quartic defocusing term, is written in physical units as \cite%
{Petrov}%
\begin{equation}
i\hbar \frac{\partial \Psi }{\partial t}=-\frac{\hbar ^{2}}{2M}\nabla
^{2}\Psi +V(\mathbf{r})\Psi +\frac{4\pi \hbar ^{2}\delta a}{M}|\Psi
|^{2}\Psi +\frac{256\sqrt{2\pi }\hbar ^{2}}{3M}a^{5/2}|\Psi |^{3}\Psi ,
\label{3DGPE}
\end{equation}%
where $\Psi $ represents equal wave functions of two components of the
binary BEC, $V(\mathbf{r})$ is the general trapping potential, $a>0$ is the
scattering length of inter-particle collisions, which induce the cubic MF
self-repulsion in each component, $\delta a\gtrless 0$, with $\left\vert
\delta a\right\vert \ll a$, represents the small disbalance of the
inter-component attraction and intra-component repulsion, the last term in
Eq. (\ref{3DGPE}) being the LHY correction to the MF equation. It is
relevant to mention that, in principle, the same equation, with $\delta a$
replaced by $a$, is valid for a single-component self-repulsive BEC.
However, without the nearly full cancellation of the MF interactions, the
LHY correction is negligibly weak, therefore it will not provide the
efficient stabilization sought for.

The reduction of Eq. (\ref{3DGPE}) to the 2D form, with coordinates $\left(
x,y\right) $, under the action of extremely tight confinement applied in the
$z$ direction, was elaborated in Ref. \cite{Petrov-Astra}, leading to an
effective cubic nonlinearity with an extra logarithmic factor,
\begin{equation}
\left( \mathrm{nonlin}\right) _{\mathrm{2D}}\sim |\Psi |^{2}\ln \left( |\Psi
|^{2}/\Psi _{0}^{2}\right) \Psi ,  \label{ln}
\end{equation}%
which is attractive and repulsive at $|\Psi |^{2}<\Psi _{0}^{2}$ and $|\Psi
|^{2}>\Psi _{0}^{2}$, respectively, where $\Psi _{0}^{2}$ corresponds to the
density determined by the equilibrium between the MF and LHY interactions,
that, in physical units, is $n_{0}=\left( 25\pi /2^{15}\right) \left( \delta
a/a\right) ^{2}a^{-3}$ \cite{Petrov}. However, this limit case corresponds
to extremely strong confinement in the $z$ direction, with the transverse
size $a_{\perp }\ll \xi $, where the healing length, corresponding to the
equilibrium density, is estimated as $\xi =\left( 32\sqrt{2}/3\pi \right)
\left( a/\left\vert \delta a\right\vert \right) ^{3/2}a\approx 5\left(
a/\left\vert \delta a\right\vert \right) ^{3/2}a$. Then, typical
experimentally relevant parameters \cite{Leticia1}-\cite{Inguscio2} yield $%
\xi \simeq 30$ nm. On the other hand, an experimentally relevant
transverse-confinement length is $a_{\perp }\sim 0.6$ $\mathrm{\mu }$m \cite%
{Leticia1}, implying relation $a_{\perp }\gg \xi $, which is opposite to the
above-mentioned necessary one. Thus, it is relevant to reduce Eq. (\ref%
{3DGPE}) to the 2D form, keeping the same nonlinearity as in Eq. (\ref{3DGPE}%
). In this connection, we note that, as straightforward analysis
demonstrates, the modified nonlinear term (\ref{ln}), corresponding to the
ultra-tight confinement, is insufficient to create a GS with a convergent
norm in 2D, yielding the density singularity $\left\vert \Psi \right\vert
^{2}\sim r^{-2}/\ln \left( r^{-1}\right) $ at $r\rightarrow 0$, hence the
two-dimensional integral (\ref{N}) is still diverging, although extremely
slowly: $N\sim \ln \left( \ln \left( r_{\mathrm{cutoff}}^{-1}\right) \right)
$, cf. Eq. (\ref{ln()}). On the other hand, the reduction of the full 3D
problem to the 2D approximation is relevant as long as the radial size of
the resulting bound state, $R$, exceeds $a_{\perp }$ by an order of
magnitude (or greater). It is expected that the predicted stable states will
have $R\sim 10$ $\mathrm{\mu }$m (see below), which justifies the latter
assumption.

To complete the derivation of the effective 2D equation, we first rescale
three-dimensional Eq. (\ref{3DGPE}), measuring the density, length, time,
and the trapping potential, respectively, in units of $\left( 36/25\right)
n_{0}$, $\xi $, $\tau \equiv \left( M/\hbar \right) \xi ^{2}$, and $\hbar
/\tau $:%
\begin{equation}
i\frac{\partial \Psi }{\partial t}=-\frac{1}{2}\nabla ^{2}\Psi +\sigma |\Psi
|^{2}\Psi +|\Psi |^{3}\Psi +V(\mathbf{r})\Psi ,  \label{3Dscaled}
\end{equation}%
where $\sigma =\pm 1$ is the sign of $\delta a$, the potential is a sum of
the above-mentioned term (\ref{U}) and the transverse-confinement one, $%
\left( 1/2\right) a_{\perp }^{-4}z^{2}$. In this notation, the
above-mentioned critical strength of the pulling potential in the linearized
version of the 3D equation (\ref{3Dscaled}) is $\left( U_{0}\right) _{%
\mathrm{coll}}=1/4$. Then, the 3D $\rightarrow $ 2D reduction is performed
by means of the standard substitution \cite{Luca,Delgado}, $\Psi \left(
x,y,z,t\right) =\psi \left( x,y,t\right) \exp \left( -z^{2}/2a_{\perp
}^{2}\right) $, followed by the averaging of Eq. (\ref{3Dscaled}) in the
transverse direction. Finally, with the help of additional rescaling, $\psi
\rightarrow \left( 2/\sqrt{5}\right) \psi $, $\left( x,y\right) \rightarrow
\left( \sqrt{5}/2\right) \left( x,y\right) $, and $t\rightarrow \left(
5/4\right) t$, the effective two-dimensional GP equation, written in terms
of the polar coordinates $\left( r,\theta \right) $ in the $\left(
x,y\right) $ plane, is cast in the form of
\begin{equation}
i\frac{\partial \psi }{\partial t}=-\frac{1}{2}\left( \frac{\partial
^{2}\psi }{\partial r^{2}}+\frac{1}{r}\frac{\partial \psi }{\partial r}+%
\frac{1}{r^{2}}\frac{\partial ^{2}\psi }{\partial \theta ^{2}}\right) -\frac{%
U_{0}}{2r^{2}}\psi +\sigma \left\vert \psi \right\vert ^{2}\psi +|\psi
|^{3}\psi ,  \label{psi2d}
\end{equation}%
which includes the pulling-to-the-center potential (\ref{U}). This equation
conserves, along with norm%
\begin{equation}
N=\int \int \left\vert \psi \left( x,y\right) \right\vert ^{2}dxdy,
\label{Npsi}
\end{equation}%
cf. Eq. (\ref{N}), the Hamiltonian,%
\begin{equation}
H=\int \int \left[ \frac{1}{2}\left( \left\vert \nabla \psi \right\vert ^{2}-%
\frac{U_{0}}{r^{2}}\left\vert \psi \right\vert ^{2}+\sigma \left\vert \psi
\right\vert ^{4}\right) +\frac{2}{5}\left\vert \psi \right\vert ^{5}\right]
dxdy,  \label{H}
\end{equation}%
and the angular momentum,%
\begin{equation}
L_{z}=i\int \int \frac{\partial \psi ^{\ast }}{\partial \theta }\psi dxdy,
\label{Lz}
\end{equation}%
where $\ast $ stands for the complex conjugation.

It is also relevant to consider the case when the balance between the
inter-component attraction and intra-component repulsion makes it possible
to set $\delta a=0$, the entire nonlinearity originating from the LHY term,
cf. Ref. \cite{LHY-only}. The respective 2D equation (produced by an
obviously different rescaling) takes the form of Eq. (\ref{psi2d}) with $%
\sigma =0$. In fact, this case is the most fundamental one, as the
suppression of the 2D collapse and formation of the GS is provided by the
quartic term, while, in any case, the cubic term plays a minor role.

The rest of the paper is organized as follows. In Section II we produce
analytical results, which are based on the asymptotic consideration of Eq. (%
\ref{psi2d}) for stationary solutions in the form of the GS and vortex
states, as well as for small perturbations which determine their stability.
Analytical considerations also make use of the Thomas-Fermi (TF)
approximation, which produces accurate results in the case of $\sigma =0$.
Numerical findings, produced by systematic simulations of Eq. (\ref{psi2d}),
are summarized in Section III. They demonstrate that the zero-vorticity
states are completely stable, in agreement with the conjecture that they
represent the system's GS, while the vortical modes feature an instability
boundary (which is predicted in an exact analytical form in Section II).
Unstable vortices (roughly speaking, those existing in the case of
relatively small $U_{0}$) spontaneously develop spiral motion of the vortex'
pivot from the center to periphery, which eventually leads to replacement of
the unstable vortex by the stable GS. A counterintuitive finding is that the
stable zero-vorticity GS can be found even in the interval of $-4/9<U_{0}<0$%
, where the central potential is (weakly) repulsive. An explanation of this
fact is given too. The paper is concluded by Section IV.

\section{Analytical considerations}

\subsection{Asymptotic forms of the solutions}

Stationary solutions to Eq. (\ref{psi2d}) with chemical potential $\mu $ and
integer vorticity (orbital quantum number) $l=0,1,2,...$, are looked for as%
\begin{equation}
\psi \left( r,t\right) =\exp \left( -i\mu t+il\theta \right) u(r),
\label{psichi2D}
\end{equation}%
with real radial function satisfying the equation%
\begin{equation}
\mu u=-\frac{1}{2}\left( \frac{d^{2}u}{dr^{2}}+\frac{1}{r}\frac{du}{dr}+%
\frac{U_{l}}{r^{2}}u\right) +\sigma u^{3}+u^{4},  \label{chi2D}
\end{equation}%
where we define%
\begin{equation}
U_{l}\equiv U_{0}-l^{2}.  \label{Ul}
\end{equation}%
The asymptotic form of the solution to Eq. (\ref{chi2D}) at $r\rightarrow 0$
is
\begin{equation}
u=\left[ \frac{1}{2}\left( U_{l}+\frac{4}{9}\right) \right]
^{1/3}r^{-2/3}-\sigma \frac{9U_{l}+4}{27U_{l}+16}+O\left( r^{2/3}\right) .
\label{r=0-2D}
\end{equation}%
The singularity of the asymptotic solution (\ref{r=0-2D}), $\sim r^{-2/3}$,
with the power which is determined by the balance between the LHY\ quartic
term and the attractive potential, and does not depend on $l$ and $\sigma $,
is weak enough to secure the convergence of the integral norm (\ref{N}) at $%
r\rightarrow 0$. It is worthy to note that, as seen in Eq. (\ref{r=0-2D}),
the expansion of the solution at $r\rightarrow 0$ is performed in powers of $%
r^{2/3}$.

Obviously, solutions to Eq. (\ref{chi2D}) may be localized at $r\rightarrow
\infty $ for $\mu <0$. Then, in the case of $\sigma =0$, a simple corollary
of Eq. (\ref{chi2D}) is an exact scaling relation which shows the dependence
of the solution on $|\mu |$:%
\begin{equation}
u\left( r;\mu \right) =\left\vert \mu \right\vert ^{1/3}u\left( |\mu
|^{-1/2}r;\mu =-1\right) .  \label{scaling}
\end{equation}%
Further, the substitution of this expression in Eq.\ (\ref{N}) yields an
exact scaling relation for the norm of the 2D state, at $\sigma =0$:%
\begin{equation}
N(\mu )=|\mu |^{-1/3}N(\mu =-1).  \label{scalingN}
\end{equation}%
We note that this relation satisfies the \textit{anti-Vakhitov-Kolokolov
(VK) criterion}, $dN/d\mu >0$, which was proposed as a necessary stability
condition for trapped modes supported by the defocusing nonlinearity \cite%
{anti} (the VK criterion proper states that $dN/d\mu <0$ is necessary for
the stability of self-trapped modes in the case of focusing nonlinearity
\cite{VK,Berge}). Below, we demonstrate that the GS family is completely
stable in the present model.

Asymptotic expression (\ref{r=0-2D}) suggests substitution
\begin{equation}
\psi \left( r,\theta ,t\right) \equiv r^{-2/3}\varphi \left( r,\theta
,t\right) ,u(r)\equiv r^{-2/3}\chi (r),  \label{uchi}
\end{equation}%
which transforms Eqs. (\ref{psi2d}) and (\ref{chi2D}) into
\begin{eqnarray}
i\frac{\partial \varphi }{\partial t} &=&-\frac{1}{2}\left[ \frac{\partial
^{2}}{\partial r^{2}}-\frac{1}{3r}\frac{\partial }{\partial r}+\frac{\left(
U_{0}+4/9\right) }{r^{2}}+\frac{1}{r^{2}}\frac{\partial ^{2}}{\partial
\theta ^{2}}\right] \varphi  \notag \\
&&+\sigma \frac{|\varphi |^{2}\varphi }{r^{4/3}}+\frac{|\varphi |^{3}\varphi
}{r^{2}},  \label{varphi}
\end{eqnarray}%
\begin{equation}
\mu \chi =-\frac{1}{2}\left[ \frac{d^{2}\chi }{dr^{2}}-\frac{1}{3r}\frac{%
d\chi }{dr}+\frac{\left( U_{l}+4/9\right) }{r^{2}}\chi \right] +\sigma \frac{%
\chi ^{3}}{r^{4/3}}+\frac{\chi ^{4}}{r^{2}}.  \label{chi}
\end{equation}%
Accordingly, the asymptotic form (\ref{chi}) of the solution at $%
r\rightarrow 0$ is replaced by a singularity-free expansion,\textbf{\ }%
\begin{equation}
\chi (r;U_{l},\sigma )=\left[ \frac{1}{2}\left( U_{l}+\frac{4}{9}\right) %
\right] ^{1/3}-\sigma \frac{9U_{l}+4}{27U_{l}+16}r^{2/3}+O\left(
r^{4/3}\right) .  \label{r=0-2D_chi}
\end{equation}

The first correction to the leading term in this expansion vanishes in the
case of $\sigma =0$ (no MF nonlinearity). Then, it follows from Eq. (\ref%
{chi}) that the expansion is replaced by
\begin{equation}
\chi (r;U_{l},\sigma =0)=\left[ \frac{1}{2}\left( U_{l}+\frac{4}{9}\right) %
\right] ^{1/3}\left( 1+\frac{2\mu }{3U_{l}}r^{2}\right) +O\left(
r^{4}\right) ,  \label{expansion}
\end{equation}%
which is valid for $U_{l}>0$. In the interval of%
\begin{equation}
-4/9<U_{l}<0  \label{-4/9}
\end{equation}
(the meaning of this interval is explained below), the quadratic term in Eq.
(\ref{expansion}) is valid too, but in that case it is not a leading
correction, as Eq. (\ref{r=0-2D_chi}) admits a stronger one, $\mathrm{const}%
\cdot r^{\beta }$, with $\beta =2/3+\sqrt{16/9+3U_{l}}<2$, while $\mathrm{%
const}$ remains indefinite, in terms of the expansion at $r\rightarrow 0$.
Exactly at $U_{l}=0$, Eq. (\ref{expansion}) is replaced by
\begin{equation}
\chi (r;U_{l}=\sigma =0)=\left( \frac{2}{9}\right) ^{1/3}\left( 1+\frac{3\mu
}{4}r^{2}\ln \left( \frac{r_{0}}{r}\right) \right) +~...,
\label{expansionU=0}
\end{equation}%
where constant $r_{0}$ is also indefinite.

The asymptotic form of the solution, given by Eq. (\ref{r=0-2D_chi}), is
meaningful if it yields $u(r)>0$ [otherwise, the derivation of Eq. (\ref%
{chi2D}) from Eq. (\ref{psi2d}) is irrelevant], i.e., for $U_{l}>0$, as well
as for \emph{weakly negative} values of the effective strength of the
central potential belonging to interval (\ref{-4/9}), which implies%
\begin{equation}
l^{2}-4/9<U_{0}\leq l^{2}.  \label{4/9}
\end{equation}%
For the vortex states, with $l\geq 1$, condition (\ref{4/9}) means, in any
case, $U_{0}>0$, but for the GS, with $l=0$, Eq. (\ref{4/9}) admits negative
$U_{0}$ with sufficiently small absolute values, \textit{viz}., $0\leq
-U_{0}<4/9$. In the limit of $U_{0}+4/9\rightarrow +0$, further analysis of
Eq. (\ref{chi}) yields the following \emph{asymptotically exact} solution,
which does not depend on $\sigma $:%
\begin{equation}
\left( \chi (r)\right) _{U_{l}+4/9\rightarrow 0}=\frac{\sqrt{3}\Gamma (1/3)}{%
\pi }\left[ \frac{|\mu |}{4}\left( U_{l}+\frac{4}{9}\right) \right]
^{1/3}r^{2/3}K_{2/3}\left( \sqrt{2|\mu |}r\right) ,  \label{exact}
\end{equation}%
where $\Gamma (1/3)\approx \allowbreak 2.68$ is the value of the
Gamma-function, and $K_{2/3}$ is the standard modified Hankel's function
(alias the modified Bessel function of the second kind). The substitution of
this expression and one defined by Eq. (\ref{uchi}) in Eq. (\ref{N}) (with $%
D=2$) yields the respective result for the norm:%
\begin{equation}
N_{U_{l}+4/9\rightarrow 0}=\frac{\Gamma ^{2}(1/3)}{\sqrt{3}}\frac{\left(
U_{l}+4/9\right) ^{2/3}}{\left( 2|\mu |\right) ^{1/3}}.  \label{exactN}
\end{equation}%
Note that this result agrees with the general scaling relation (\ref%
{scalingN}).

While the existence of the bound state under the combined action of the
repulsive potential and dominating defocusing quartic nonlinearity is a
counterintuitive finding, it is closely related to the previously known fact
that the 2D nonlinear Schr\"{o}dinger equations with the repulsive nonlinear
term $|u|^{p-1}u$ supports localized solutions with a singular asymptotic
form,%
\begin{equation}
u\approx \left[ \frac{1}{2}\left( \frac{4}{\left( p-1\right) ^{2}}%
+U_{0}\right) \right] ^{1/\left( p-1\right) }r^{-2/\left( p-1\right) }
\label{p}
\end{equation}%
at $r\rightarrow 0$ \cite{Veron} [here, Eq. (\ref{p}) takes into account the
presence of the potential with strength $U_{0}$ in Eq. (\ref{chi2D}), which
was not included in Ref. \cite{Veron}]. It is seen that this singular state
exists at $0<-U_{0}<4/\left( p-1\right) ^{2}$, and the integral which
defines its norm converges at $r\rightarrow 0$ for $p>3$, including the case
of the quartic nonlinearity, with $p=4$. This result, which was originally
obtained as a formal one \cite{Veron}, may be understood, in terms of the
physical realization, as an effect created by an additional delta-functional
\emph{attractive} potential, which is concentrated on a sphere of a
vanishingly small radius $\rho $:%
\begin{equation}
U_{\delta }=-\lim_{\rho \rightarrow 0}\left[ \varepsilon \delta \left(
r-\rho \right) \right] ,~\varepsilon \equiv \left( p-1\right) ^{-1}\rho
^{-1}.  \label{delta}
\end{equation}%
This potential, added to the model, becomes \textquotedblleft invisible" in
the limit of $\rho \rightarrow 0$ in Eq. (\ref{delta}), being screened by
the singular profile of the pinned state \cite{HS}. This consideration
explains the possibility of the existence of the bound state which is not
supported by any apparent factor pulling the wave function to the center.
Note that the \textquotedblleft charge" of the invisible potential, $Q\equiv
2\pi \rho \cdot \varepsilon =2\pi /\left( p-1\right) $, remains finite in
the limit of $\delta \rightarrow 0$. In similar 1D and 3D settings, the
screened charge is, respectively, diverging or vanishingly small, $Q_{%
\mathrm{1D}}\sim \rho ^{-1}$ and $Q_{\mathrm{3D}}\sim \rho $ \cite{HS}, see
also a brief review of the topic in Ref. \cite{we}.

In the limit of $r\rightarrow \infty $, the asymptotic form of the solution
to Eq. (\ref{chi}) is%
\begin{equation}
\chi (r)\approx \chi _{0}r^{1/6}\exp \left( -\sqrt{2|\mu |}r\right) ,
\label{chi0}
\end{equation}%
where $\chi _{0}$ is an arbitrary constant (in terms of the asymptotic form
at $r\rightarrow \infty $), and $\mu $ must be negative. An approximate
global interpolation for the solution may be produced by combining
asymptotic forms (\ref{r=0-2D_chi}) and (\ref{chi0}):%
\begin{equation}
\chi _{\mathrm{interpol}}(r)=\left[ \frac{1}{2}\left( U_{l}+\frac{4}{9}%
\right) \right] ^{1/3}\exp \left( -\sqrt{2|\mu |}r\right) .  \label{inter}
\end{equation}%
Of course, this is a coarse approximation, as it postulates a wrong power of
the pre-exponential factor at $r\rightarrow \infty $ [$0$ instead of $1/6$,
see Eq. (\ref{chi0})], and ignores any effect of term $u^{3}$ in Eq. (\ref%
{chi2D}). The calculation of norm (\ref{N}) with the approximate solution
given by Eq. (\ref{inter}) produces the respective analytical expression,%
\begin{equation}
N_{\mathrm{interpol}}(\mu )=2\pi \int_{0}^{\infty }u_{\mathrm{interpol}%
}^{2}(r)rdr=\pi \Gamma \left( 2/3\right) \frac{\left( U_{l}+4/9\right) ^{2/3}%
}{\left( 4|\mu |\right) ^{1/3}},  \label{Ninter}
\end{equation}%
[where $\Gamma (2/3)\approx \allowbreak 1.354$ is the value of the
Gamma-function], which also agrees with scaling (\ref{scalingN}). In the
limit of $U_{l}+4/9\rightarrow 0$, the difference between the approximate
value of the norm, given by Eq. (\ref{Ninter}), and the asymptotically exact
result (\ref{exactN}) amounts to a factor $\approx \allowbreak 0.81$.

\subsection{Vortex states and their stability}

Usually, the presence of integer vorticity $l\geq 1$ implies that the
amplitude vanishes at $r\rightarrow 0$ as $r^{l}$, which is necessary
because the phase of the vortex field is not defined at $r=0$. However, the
indefiniteness of the phase is also compatible with the amplitude diverging
at $r\rightarrow 0$. In the linear equation, this divergence has the
asymptotic form of the standard Neumann's (alias singular Bessel's)
cylindrical function, $Y_{l}(r)\sim r^{-l}$, which makes the respective 2D
state unnormalizable (i.e., physically irrelevant) for all values $l\geq 1$.
However, in the present system Eq. (\ref{r=0-2D}) demonstrates that the
interplay of the central potential and quartic nonlinearity reduces the
divergence to the level of $r^{-2/3}$, for any $l$, thus maintaining the
normalizability of the states under the consideration, similar to what was
found in Ref. \cite{HS1}, where the quintic repulsive nonlinearity kept the
singularity in another form which secured the convergence of the 2D integral
norm (\ref{N}), $u(r)\sim r^{-1/2}$.

Stationary solutions for vortex modes, as produced by Eq. (\ref{chi}), are
not essentially different from the GS ones, as the presence of vorticity $l$
affects only the effective potential strength defined by Eq. (\ref{Ul}).
Real difference between the states with $l=0$ and $l\geq 1$ is revealed by
the analysis of their stability. To this end, it is necessary to derive the
linearized Bogoliubov - de Gennes (BdG) equations for eigenmodes of small
perturbations, with arbitrary integer azimuthal index $m$ and instability
growth rate $\lambda $ (which may be complex), added to the stationary
states. It is natural to perform this analysis in terms of Eq. (\ref{varphi}%
), which makes it possible to eliminate the singular factor, $r^{-2/3}$,
from the perturbations. Thus, the perturbed solution is looked for in the
usual form \cite{Yang},%
\begin{gather}
\varphi (r,\theta ,t)=\exp \left( -i\mu t+il\theta \right)   \notag \\
\times \left[ \chi (r)+v_{1}(r)\exp \left( \lambda t+im\theta \right)
+v_{2}^{\ast }(r)\exp \left( \lambda ^{\ast }t-im\theta \right) \right] ,
\label{chi_perturbed}
\end{gather}%
where $\chi (r)$ is a solution of Eq. (\ref{chi}). The substitution of this
in Eq. (\ref{psi2d}) and linearization with respect to perturbation
amplitudes $v_{1,2}$ leads to BdG equations in the radial form:
\begin{eqnarray}
i\lambda v_{1} &=&-\frac{1}{2}\left[ \frac{d^{2}}{dr^{2}}-\frac{1}{3r}\frac{d%
}{dr}+\frac{U_{0}-(l+m)^{2}+4/9}{r^{2}}\right] v_{1}  \notag \\
&&+\sigma \frac{\chi _{0}^{2}}{r^{4/3}}(2v_{1}+v_{2})+\frac{\chi _{0}^{3}}{%
2r^{2}}(5v_{1}+3v_{2}),  \notag \\
-i\lambda v_{2} &=&-\frac{1}{2}\left[ \frac{d^{2}}{dr^{2}}-\frac{1}{3r}\frac{%
d}{dr}+\frac{U_{0}-(l-m)^{2}+4/9}{r^{2}}\right] v_{2}  \notag \\
&&+\sigma \frac{\chi _{0}^{2}}{r^{4/3}}(2v_{2}+v_{1})+\frac{\chi _{0}^{3}}{%
2r^{2}}(5v_{2}+3v_{1}).  \label{chi_BdG}
\end{eqnarray}%
The instability driven by the perturbation eigenmode with $m\geq 2$ splits
the vortex in $m$ fragments, while the eigenmode with $m=1$ is a dipole
perturbation which drives spontaneous drift of the vortex' pivot from the
original position \cite{old-review}. By solving the eigenvalue problem based
on Eq. (\ref{chi_BdG}), one can find the spectrum of instability growth
rates $\lambda $, and thus distinguish stable solution as those for which
all the eigenvalues are imaginary.

It is relevant to consider the form of eigenmodes produced by Eqs. (\ref%
{chi_BdG}) at $r\rightarrow 0$. In this limit, solutions are looked for as
\begin{equation}
v_{1,2}(r)=v_{1,2}^{(0)}r^{\gamma },  \label{gamma}
\end{equation}%
where $\gamma $ may be complex, and $v_{1,2}^{(0)}$ are constants. Relevant
eigenmodes may not be singular at $r\rightarrow 0$, as a singular mode,
assuming with very large local values, is incompatible with the
linearization procedure. Thus, relevant are values of $\gamma $ with \textrm{%
Re}$(\gamma )>0$.

On the substitution of expression (\ref{gamma}) in Eq. (\ref{chi_BdG}), the
condition of the cancellation of singular terms $\sim r^{-2}$ leads to the
following quadratic equations for $\gamma $ (either of them must hold):%
\begin{equation}
\gamma ^{2}-\frac{4}{3}\gamma -3\chi ^{3}(r=0)-m^{2}\mp \sqrt{%
4l^{2}m^{2}+9\chi ^{6}(r=0)}=0,  \label{quadr}
\end{equation}%
where $\chi (r=0)$ is given by Eq. (\ref{r=0-2D_chi}):%
\begin{equation}
\chi ^{3}(r=0)=\frac{1}{2}\left( U_{0}-l^{2}+\frac{4}{9}\right) ,
\label{chi00}
\end{equation}%
[recall this expression is relevant if it yields $\chi ^{3}(r=0)>0$, i.e., $%
U_{0}>l^{2}-4/9$], and Eq. (\ref{chi00}) is used to eliminate $U_{0}$ from
Eq. (\ref{quadr}). Note that, in the lowest asymptotic approximation, the
solution at $r\rightarrow 0$ is not affected by terms $\sim \sigma $ in Eq. (%
\ref{chi_BdG}).

In the case of $l=0$ (the GS with no vorticity), Eq. (\ref{quadr})
simplifies to the following pair of equations:
\begin{equation}
\gamma ^{2}-\frac{4}{3}\gamma -6\chi ^{3}(r=0)-m^{2}=0,  \label{minus}
\end{equation}%
\begin{equation}
\gamma ^{2}-\frac{4}{3}\gamma -m^{2}=0,  \label{plus}
\end{equation}%
It is obvious that each equation (\ref{minus}) and (\ref{plus}) produces one
root $\gamma >0$ and one $\gamma <0$, only the former one being relevant, as
said above. In the case of the underlying vortex state, with $l^{2}\geq 1$,
Eq. (\ref{quadr}) with the top sign in front of the square root leads to the
same conclusion. On the other hand, Eq. (\ref{quadr}) with the bottom sign
gives rise to two relevant roots (instead of the single one), with \textrm{Re%
}$(\gamma )>0$, when the free term in the corresponding quadratic equation (%
\ref{quadr}) for $\gamma $ is positive, i.e.,
\begin{equation}
3\chi ^{3}(r=0)+m^{2}<\sqrt{4l^{2}m^{2}+9\chi ^{6}(r=0)}.  \label{<}
\end{equation}%
Further, the substitution of expression (\ref{chi00}) in Eq. (\ref{<}) leads
to the following condition:
\begin{equation}
U_{0}<\left( U_{0}\right) _{\mathrm{crit}}=\frac{1}{3}\left( 7l^{2}-m^{2}-%
\frac{4}{3}\right) .  \label{U0<}
\end{equation}

Equation (\ref{U0<}) never holds for the GS, with $l=0$. On the other hand,
for $l^{2}\geq 1$, the largest area in which Eq. (\ref{U0<}) holds
corresponds to $m=\pm 1$ (the eigenmode of the drift perturbation):%
\begin{equation}
U_{0}<\left( U_{0}\right) _{\mathrm{crit}}=(7/9)\left( 3l^{2}-1\right) .
\label{lstability}
\end{equation}%
Finally, for the practically important case of $l=1$, which we consider
below, Eq. (\ref{lstability}) reduces to%
\begin{equation}
U_{0}<\left( U_{0}\right) _{\mathrm{crit}}^{(l=1)}=14/9.  \label{14/9}
\end{equation}
Note also that Eq. (\ref{U0<}) formally holds for all $l^{2}\geq 1$ and $m=0$%
. However, the above derivation is irrelevant for $m=0$.

Thus, if condition (\ref{U0<}) holds, Eq. (\ref{chi_BdG}) gives rise to
additional eigenmodes whose eigenvalues may (or may not) be unstable. As
shown in the following section, numerical solution of the BdG equations (\ref%
{chi_BdG}), confirmed by direct simulations of perturbed evolution of the
vortex modes in the framework of Eq. (\ref{varphi}), corroborates the
conjecture that, in the region defined by Eq. (\ref{14/9}), the vortices
with $l=1$ are unstable against spontaneous onset of the outward drift of
the vortex' pivot, and ones with $l=2$ are unstable too in the region
defined by Eq. (\ref{lstability}) with $l=2$. Up to the accuracy of the
numerical data, $U_{0}=14/9$ is indeed identified as the stability boundary
for the vortex modes with $l=1$.

\subsection{The Thomas-Fermi (TF) approximation}

Another analytical method is offered by the TF approximation, which amounts
to dropping the derivatives in Eq. (\ref{chi}), assuming $U_{0}\gg 1$,
irrespective of the value of $|\mu |$ (in fact, it is shown below that the
approximation may produce relevant results even when $U_{0}$ is not very
large). This simplification yields an explicit approximate solution in the
case of $\sigma =0$ (if the nonlinearity is furnished solely by the LHY
term):%
\begin{equation}
\chi _{\mathrm{TF}}(r)=\left\{
\begin{array}{c}
\left[ \left( U_{l}+4/9\right) /2-|\mu |r^{2}\right] ^{1/3},~\mathrm{at}%
~~r<r_{0}\equiv \sqrt{\left( U_{l}+4/9\right) /\left( 2|\mu |\right) }, \\
0,~\mathrm{at}~~r>r_{0}~,%
\end{array}%
\right.  \label{TF}
\end{equation}%
for $\mu <0$. In the limit of $r\rightarrow 0$, Eq. (\ref{TF}) yields the
same exact value of $\chi (r=0)=\left[ \left( U_{l}+4/9\right) /2\right]
^{1/3}$ as given by Eq. (\ref{r=0-2D}). On the other hand, the TF\
approximation predicts a finite radius $r_{0}$ of the GS, neglecting the
exponentially decaying tail at $r\rightarrow \infty $, cf. Eq. (\ref{chi0}).

Further, the TF approximation given by Eq. (\ref{TF}) makes it possible to
calculate the corresponding $N(\mu )$ dependence for the GS family:%
\begin{equation}
N_{\mathrm{TF}}^{(\sigma =0)}(\mu )=2\pi \int_{0}^{r_{0}}\left[ r^{-2/3}\chi
_{\mathrm{TF}}(r)\right] ^{2}rdr=C\frac{U_{l}+4/9}{|\mu |^{1/3}},
\label{NTF}
\end{equation}%
where a numerical constant is $C\equiv \pi \int_{0}^{1}\left(
x^{-2}-1\right) ^{2/3}xdx\approx 3.80$, cf. Eq. (\ref{Ninter}). Note that,
similar to Eq. (\ref{Ninter}), this result complies with Eq. (\ref{scalingN}%
), and at $U_{l}=0$ the comparison of approximate values given by Eqs. (\ref%
{Ninter}) and (\ref{NTF}) is $\left\{ N_{\mathrm{interpol}}(\mu )/N_{\mathrm{%
TF}}^{(\sigma =0)}(\mu )\right\} |_{U_{l}=0}\approx \allowbreak 0.92$.

It is relevant to mention that TF radius $r_{0}$ keeps the same value, as
given by Eq. (\ref{TF}), in the presence of the MF defocusing cubic term
with $\sigma =1$ in Eq. (\ref{chi2D}), although the shape of the GS is more
complex. In this case, the asymptotic form of the respective $N_{\mathrm{TF}%
}^{(\sigma =1)}(\mu )$ dependence at $\mu \rightarrow -\infty $ is the same
as given by Eq. (\ref{NTF}), while in the limit of $\mu \rightarrow -0$ the
analysis produces a different result, with a much weaker singularity:%
\begin{equation}
N_{\mathrm{TF}}^{(\sigma =1)}(\mu )\approx \left( \pi /2\right) U_{l}\ln
\left( 1/|\mu |\right) .  \label{Uln}
\end{equation}%
Although this dependence $N(\mu )$ is different from that given by Eq. (\ref%
{scalingN}) in the absence of the cubic term ($\sigma =0$), Eq. (\ref{Uln})
also agrees with the anti-VK criterion.

Even in the case of the \emph{focusing sign} of the MF term, corresponding
to $\sigma =-1$ in Eq. (\ref{chi2D}), the LHY-induced quartic nonlinearity
is able to stabilize the condensate against the combined action of the MF
self-attraction and pull to the center. In this case, the TF approximation,
applied to Eq. (\ref{chi2D}), cannot be easily resolved to predict $u_{%
\mathrm{TF}}(r)$, but it produces an inverse dependence, for $r$ as a
function of $u$:%
\begin{equation}
r^{2}=\left( U_{l}/2\right) \left( -\mu -u^{2}+u^{3}\right) ^{-1}  \label{r}
\end{equation}%
[Eq. (\ref{chi2D}) is easier to use for this purpose than Eq. (\ref{chi})].
Then, looking for a maximum of expression (\ref{r}), which is attained at $%
u_{\max }=2/3$, it is easy to find the corresponding size of the TF state:
\begin{equation}
r_{0}^{(\sigma =-1)}=\frac{U_{l}}{2\left( |\mu |-4/27\right) },  \label{-1}
\end{equation}%
which, in turn, suggests that the GS exists\ in the case of $\sigma =-1$,
provided that $|\mu |$ exceeds a threshold value,
\begin{equation}
|\mu |>\left( |\mu |\right) _{\mathrm{thr}}=4/27.  \label{thr}
\end{equation}%
Pursuant to Eq. (\ref{-1}), the norm diverges at $\mu \rightarrow -4/27$ as%
\begin{equation}
N\approx 2\pi \left( r_{0}^{(\sigma =-1)}\right) ^{2}u_{\max }^{2}=\frac{%
2\pi }{9}\frac{U_{l}^{2}}{\left( |\mu |-4/27\right) ^{2}},  \label{NN}
\end{equation}

Finally, we note that, in the absence of any potential ($U_{0}=0$), the
interplay of the cubic self-attraction and quartic repulsion may readily
create stable multidimensional solitons, including ones with embedded
vorticity $l=1$ \cite{Barcelona}. The analytical predictions obtained in
this section are compared below to their numerically found counterparts in
Figs. \ref{fig1} - \ref{fig4}.

\section{Numerical results}

\subsection{Stable ground-state (GS) solutions}

Stationary solutions of Eq. (\ref{chi}) were produced by means of the
Newton's iteration method. Stability of stationary solutions was identified
by means of the numerically solved linearized eigenvalue problem for small
perturbations, based on Eq. (\ref{chi_BdG}). As said above, the stability
condition is that all eigenvalues $\lambda $ must have zero real parts.
Then, the so predicted (in)stability was verified by direct simulations of
underlying equation (\ref{psi2d}). The simulations were run by means of the
split-step Fourier-transform method \cite{Yang}, implemented with the help
of the Runge-Kutta numerical scheme. An absorber, installed at edges of the
integration domain, was employed to prevent reflection of the emitted
radiation, without affecting the mode under the consideration. To this end,
the size of the domain was always taken to be much larger than the mode's
scale, and it was checked that the results were not affected by the size.
The analysis was performed for the model including the cubic self-defocusing
or focusing term, i.e., with $\sigma =\pm 1$ in Eq. (\ref{psi2d}), as well
as for the most fundamental case of $\sigma =0$, when the nonlinearity is
provided solely by the LHY effect.

\begin{figure}[tbp]
\subfigure[]{\includegraphics[width=3.2in]{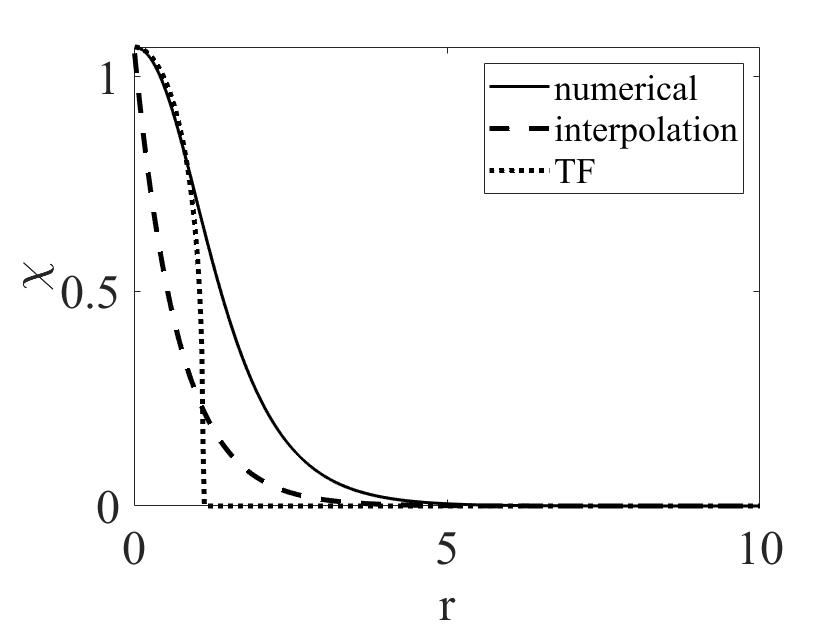}}%
\subfigure[]{
\includegraphics[width=3.2in]{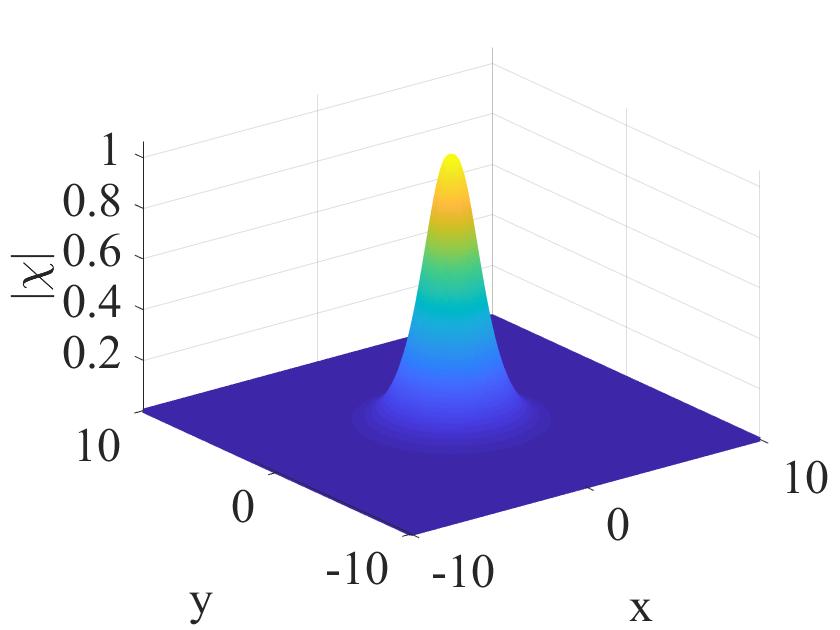}} 
\caption{(a) The radial profile of a (stable) numerically found GS and its
TF and interpolation counterparts, produced by Eqs. (\protect\ref{chi}), (%
\protect\ref{TF}), and (\protect\ref{inter}), respectively, with $\protect%
\sigma =0$, $U_{0}=2$, and $\protect\mu =-1$. The total norm of the
numerical, TF, and interpolation solutions are, severally, $N_{\mathrm{num}%
}=10.34$, $N_{\mathrm{TF}}=9.28$, and $N_{\mathrm{interpol}}=4.86$. (b) The
global shape of the numerically generated solution.}
\label{fig1}
\end{figure}

First, Fig. \ref{fig1} displays a typical example of the stable GS, obtained
as a numerical solution of Eq. (\ref{chi}) with $\sigma =0$, $U_{0}=2$, and $%
|\mu |=1$, along with its analytical counterparts, produced by the TF
approximation and by the interpolating approximation (\ref{inter}), for the
same $\mu $, as per Eqs. (\ref{TF}) and (\ref{inter}), respectively.\ It is
seen that the approximations are not accurate in this case. In particular,
the TF solution is relatively close to the numerical counterpart only in its
central core, because condition $U_{0}\gg 1$ does not hold in this case. The
discrepancy in the total norm, calculated as per Eqs. (\ref{N}) and (\ref%
{NTF}) for the numerically exact and TF solutions, is $\left( N-N_{\mathrm{TF%
}}\right) /N\approx 10\%$ [it is smaller than it may seem in Fig. \ref{fig1}%
(a) because relation (\ref{uchi}) suppresses the contribution of the region
of larger $r$, where the TF approximation is wrong].

\begin{figure}[tbp]
\subfigure[]{\includegraphics[width=3.2in]{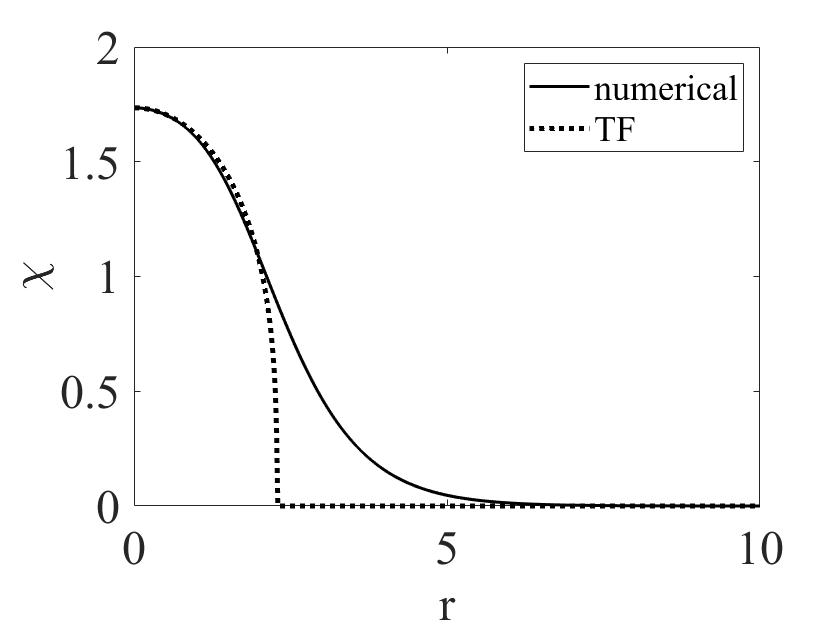}}\subfigure[]{%
\includegraphics[width=3.2in]{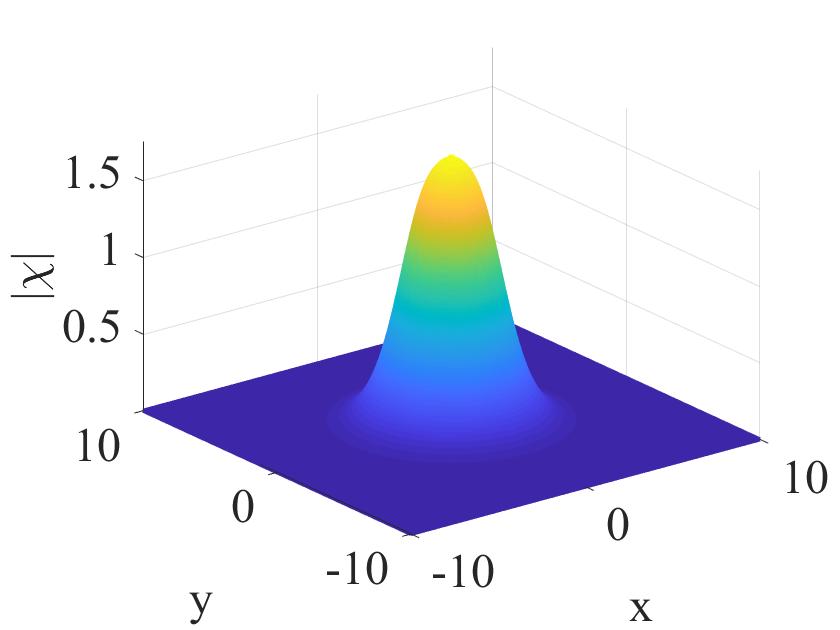}}
\caption{The same as in Fig. \protect\ref{fig1}, but for $U_{0}=10$ and
without showing the interpolation given by Eq. (\protect\ref{inter}), which
is irrelevant in this case. The norms of the numerical and approximate
solutions are $N_{\mathrm{num}}=41.05$, $N_{\mathrm{TF}}=39.68$.}
\label{fig2}
\end{figure}

\begin{figure}[tbp]
{\includegraphics[width=3.2in]{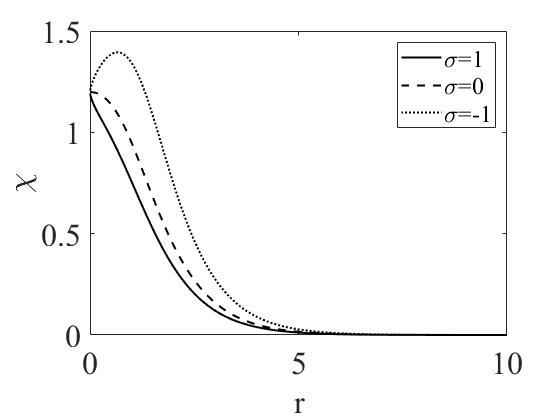}}
\caption{Examples of numerically found stable GS solutions of Eq. (\protect
\ref{chi}) for $l=0,U_{0}=3,\protect\mu =-0.8$ , and $\protect\sigma =1$, $0$%
, and $-1$. The respective norms are $N(\protect\sigma =1)=11.7$, $N(\protect%
\sigma =0)=15.41$ and $N(\protect\sigma =-1)=24.62$.}
\label{fig7}
\end{figure}

Further, Fig. \ref{fig2} displays the GS, along with the corresponding TF
approximation, for sufficiently large $U_{0}=10$. It is seen that, as
expected, the TF solution is virtually identical to the numerical one at $%
r<r_{0}\approx \allowbreak 2.\,\allowbreak 24$, see Eq. (\ref{TF}), while
the decaying tail is ignored by TF. In this case, the discrepancy in the
total norm is $\left( N-N_{\mathrm{TF}}\right) /N\approx 3.3\%$.

The effect of the MF cubic term of either sign, repulsive ($\sigma =1$) or
attractive ($\sigma =-1$) on the shape of the GS is illustrated by Fig. \ref%
{fig7}. At $r=0$, all the three shapes converge to a common value, $\chi
(r=0)\approx 1.20$, exactly as predicted by Eq. (\ref{r=0-2D_chi}).

\begin{figure}[tbp]
\subfigure[]{\includegraphics[width=3.2in]{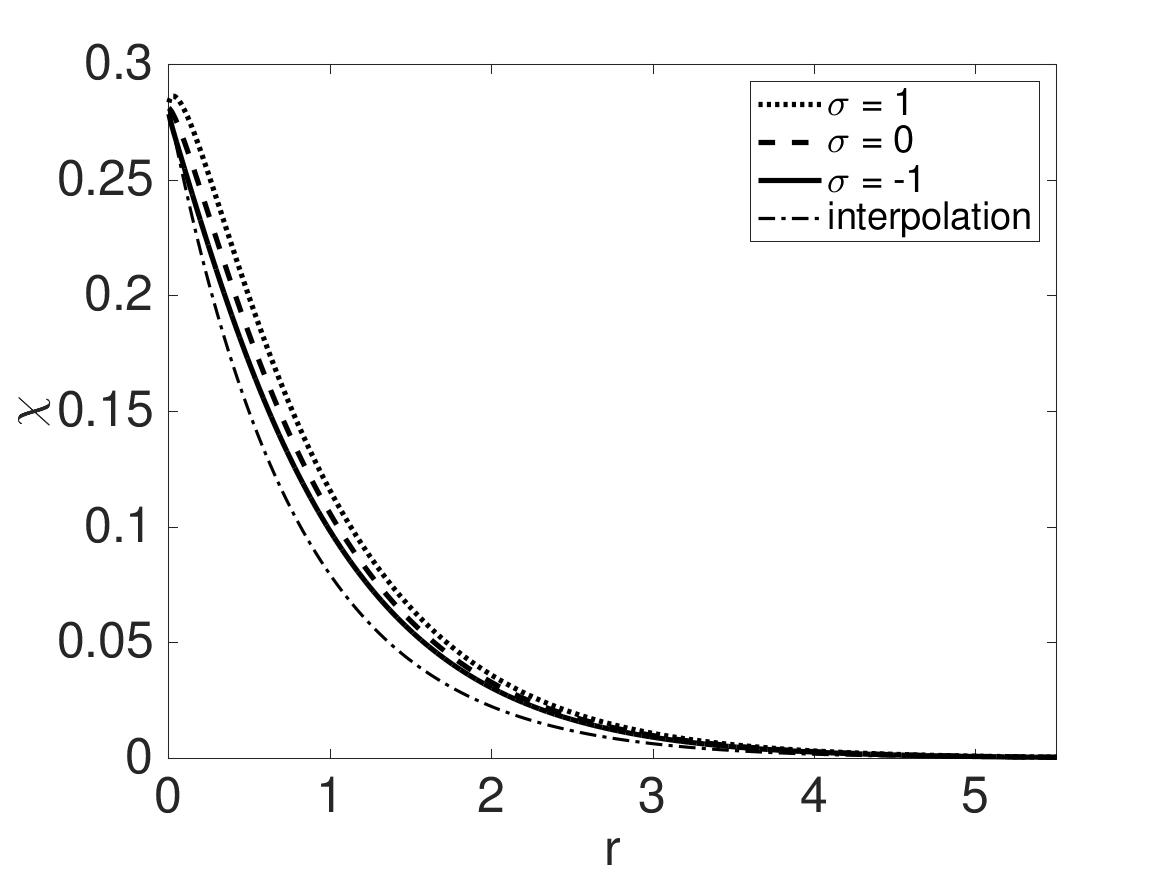}}\subfigure[]{%
\includegraphics[width=3.2in]{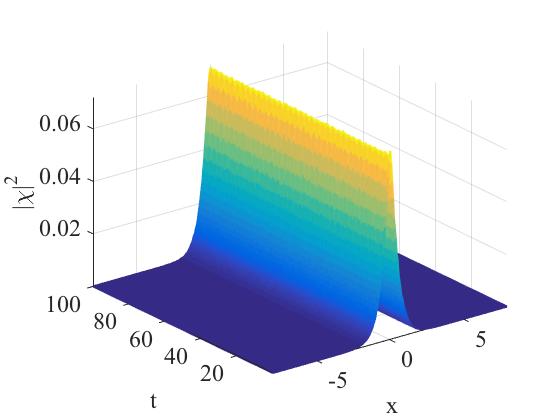}}
\caption{(a) Radial profiles of stable GS modes, produced by the numerical
solution of Eq. (\protect\ref{chi}), and by interpolation (\protect\ref%
{inter}), in the presence of the \emph{repulsive potential}, with $U_{0}=-0.4
$, at $\protect\mu =-0.8$, without and with the MF repulsive or attractive
cubic term ($\protect\sigma =0$ and $\protect\sigma =1$ or $-1$,
respectively). The corresponding values of the norm are $N\left( \protect%
\sigma =1\right) =0.41$, $N(\protect\sigma =0)=0.45$, and $N(\protect\sigma %
=-1)=0.52$, the interpolating approximation giving $N\approx 0.36$, pursuant
to Eq. (\protect\ref{Ninter}). (b) Confirmation of the stability of the GS
mode in direct simulations, in the case of $\protect\sigma =1$. }
\label{fig3}
\end{figure}

The above counterintuitive prediction of the GS solutions existing in the
presence of the \emph{repulsive potential}, with $U_{0}$ belonging to
interval (\ref{-4/9}), is confirmed by numerical results. As an example,
Fig. \ref{fig3} displays stable GSs which were found, in the numerical form,
at $U_{0}=-0.4$, taken close to the edge of the interval, $U_{0}=-4/9$, for
all the three essential values, $\sigma =0$ and $\pm 1$, of the coefficient
in front of the MF cubic repulsive term. The figure shows that the
interpolating approximation, generated by Eq. (\ref{inter}), is quite
accurate in this case, while the MF cubic term produces a weak effect on the
solution. As for the TF approximation (\ref{TF}), it does not apply to $%
U_{0}<0$.

\begin{figure}[tbp]
\subfigure[]{\includegraphics[width=3.2in]{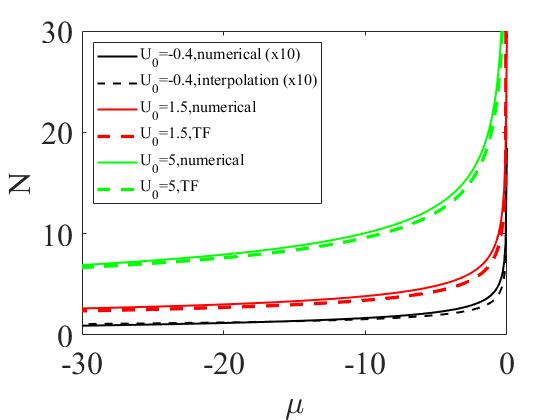}} \subfigure[]{%
\includegraphics[width=3.2in]{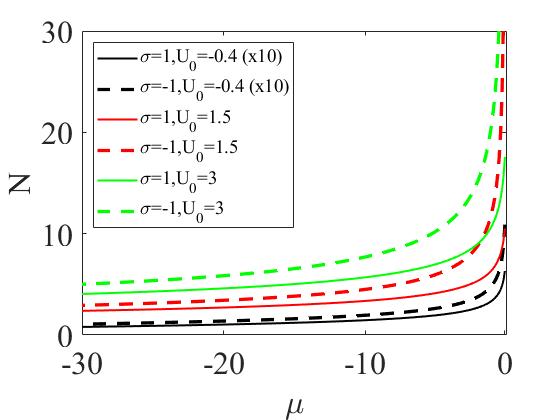}} \subfigure[]{%
\includegraphics[width=3.2in]{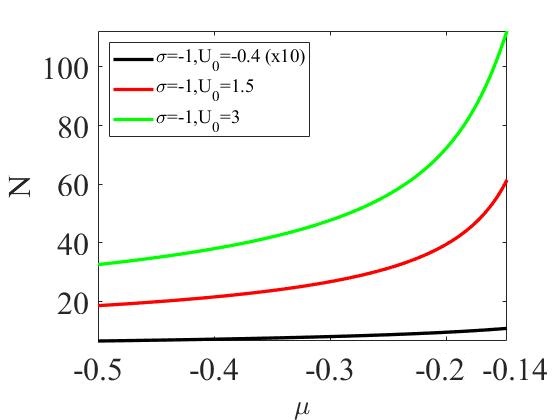}}
\caption{Dependences $N(\protect\mu )$ for stable GS solutions with $%
U_{0}=-0.4$, and $N(\protect\mu )$ for $U_{0}=$ $1.5$, $3.0$, $5.0$, which
correspond, respectively, to the repulsive and attractive central potential.
To plot the curve for $U_{0}=-0.4$, values of $N$ are multiplied by $10$, as
the actual values of the norm are too small in this case. Panels (a) and (b)
correspond to the system which does or does not include the repulsive or
attractive cubic term ($\protect\sigma =0$ and $1,-1$, respectively). In
(a), the numerical results are compared to the analytical ones, predicted by
the TF approximation, as per Eq. (\protect\ref{NTF}) [except for the case of
$U_{0}=-0.4$, where the TF approximation is irrelevant; however, in this
case the analytical prediction given by the interpolating approximation, in
the form of Eq. (\protect\ref{Ninter}), completely overlaps with the
numerically generated curve]. In (b), the numerically generated curves are
compared for $\protect\sigma =1$ and $-1$. Panel (c) displays a zoom of the
plot from (b) at small values of $|\protect\mu |$, with the aim to show the
proximity to the threshold value $\left( |\protect\mu |\right) _{\mathrm{thr}%
}=4/27\approx \allowbreak 0.15$ for $\protect\sigma =-1$, as predicted in
the TF approximation by Eq. (\protect\ref{thr}) (see further comments,
concerning this point, in the text).}
\label{fig4}
\end{figure}

Families of the GS solutions are characterized by the corresponding
dependences $N(\mu )$. These results are summarized in Figs. \ref{fig4}(a,b)
for the 2D system without and with the MF repulsive or attractive cubic
term, $\sigma =0$ and $\sigma =\pm 1$, and for several values of strength $%
U_{0}$ of the central potential (including both $U_{0}>0$ and $U_{0}<0$). In
particular, the MF cubic term produces a more considerable effect with the
increase of $U_{0}$, which is naturally explained by the fact that the
solution's amplitude is larger for larger $U_{0}$, as per Eqs. (\ref%
{r=0-2D_chi}), (\ref{inter}), and (\ref{TF}). The $N(\mu )$ curves produced
by the TF approximation pursuant to Eq. (\ref{NTF}) are compared to their
numerical counterparts in panel (a). As mentioned above, the accuracy of the
TF approximation essentially improves with the increase of $U_{0}$. On the
other hand, interpolating approximation produces poor accuracy at $U_{0}>0$,
but for $U_{0}=-0.4$ this approximation is virtually identical to the
numerical counterpart.

Panel (c) in Fig. \ref{fig4} confirms that, in the presence of the
attractive cubic term ($\sigma =-1$), the GS exists at $|\mu |>\left( |\mu
|\right) _{\mathrm{thr}}$, as predicted by the TF approximation in Eq. (\ref%
{thr}). Up to the accuracy of the numerical results, the threshold value of $%
|\mu |$ is indeed $4/27$, in agreement with Eq. (\ref{thr}). This finding is
explained by the fact that, as it follows from Eq. (\ref{-1}), the width of
the GS diverges in the limit of $|\mu |\rightarrow \left( |\mu |\right) _{%
\mathrm{thr}}$, hence in this limit the spatial derivatives in Eq. (\ref%
{chi2D}) become negligible, and the TF approximation becomes asymptotically
exact. Strictly speaking, $N(\mu )$ steeply diverges in the limit of $|\mu
|\rightarrow 4/27$, according to Eq. (\ref{NN}), but it is difficult to plot
the curves very close to the threshold, as the bound states become extremely
broad in this limit.

Concerning the stability, both the computation of eigenvalues for small
perturbations and direct simulations of perturbed evolution demonstrate
\emph{complete stability} of the fundamental (GS) solutions at all values of
$U_{0}$, both positive ones and negative values belonging to interval (\ref%
{-4/9}), and at all (negative) values of $\mu $. As an illustration, Fig. %
\ref{fig3}(b) demonstrates the stability of the GS in the counterintuitive
case of the repulsive central potential, with $U_{0}=-0.4$. The stability
does not depend either on the presence of the MF cubic term, being equally
valid for $\sigma =0$ and $\pm 1$. Note also that the anti-VK stability
criterion, $dN/d\mu >0$, holds for all $N(\mu )$ curves in Fig. \ref{fig4}.

\subsection{Vortex modes}

\begin{figure}[tbp]
\subfigure[]{\includegraphics[width=3.2in]{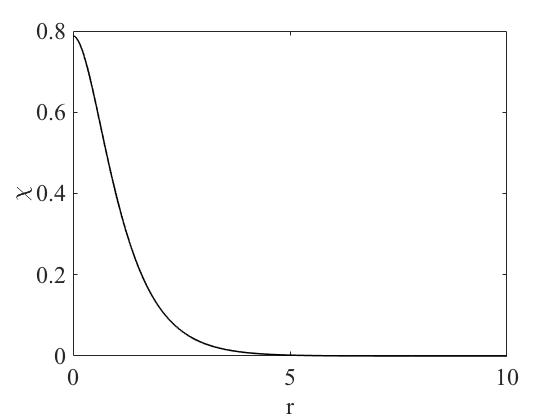}}\subfigure[]{%
\includegraphics[width=3.2in]{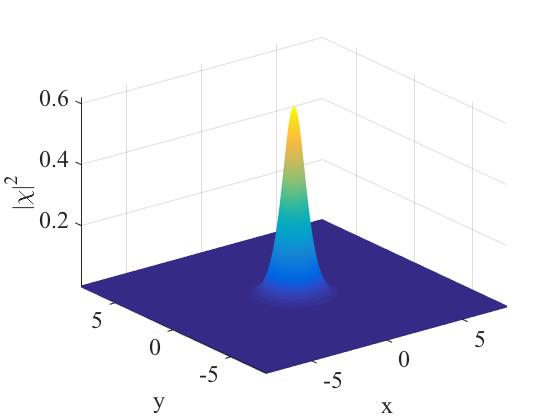}}
\caption{The amplitude structure of an (unstable) vortex mode with $l=1$ in
Eq. (\protect\ref{psichi2D}), numerically generated for $\protect\sigma =0$,
$U_{0}=1.53$ and $\protect\mu =-1$: (a) the radial profile; (b) the global
shape. The total norm of the vortex mode, computed as per Eq. (\protect\ref%
{uchi}), is $N\approx 4.58$. }
\label{fig5}
\end{figure}

\begin{figure}[tbp]
\subfigure[] {\includegraphics[width=3.2in]{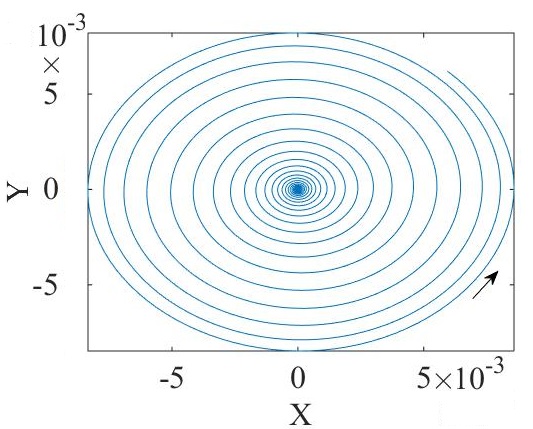}} \subfigure[] {%
\includegraphics[width=3.2in]{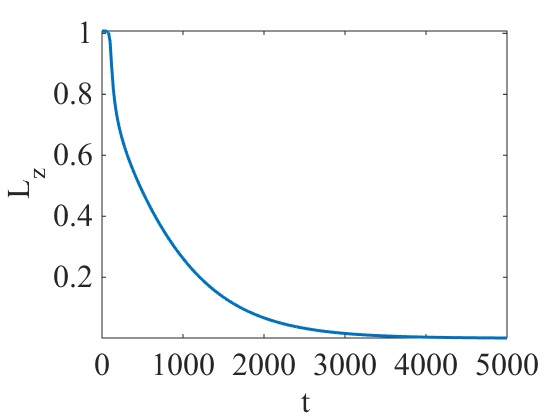}} 
\caption{The instability development of the vortex mode with $l=1$ which is
shown in Fig. \protect\ref{fig5}. (a) The trajectory of spontaneous motion
of the vortex' CM [see Eq. (\protect\ref{cm})], initiated by the drift
instability, at the initial stage of the evolution. The arrow indicates the
direction of the motion along the trajectory. (b) The evolution of the total
angular momentum, defined as per Eq. (\protect\ref{Lz}), gradually consumed
by the edge absorber, illustrates the spontaneous transformation of the
unstable vortex into a stable ground state.}
\label{fig6}
\end{figure}

As mentioned above, the stationary shape of vortex modes, given by Eq. (\ref%
{psichi2D}) with $l\geq 1$, is actually the same as for the GS with $l=0$,
the difference amounting to replacement of $U_{0}$ by $U_{l}~$as per Eq. (%
\ref{Ul}). A typical example of the vortex solution with $U_{0}=1.53$ and $%
l=1$ is displayed in Fig. \ref{fig5} [as it follows from Eqs. (\ref{chi})
and (\ref{Ul}), its amplitude profile is the same as that of the GS with $%
U_{0}=0.53$]. This value of $U_{0}$ is chosen in the instability region,
close to its boundary predicted by Eq. (\ref{14/9}), $\left( U_{0}\right) _{%
\mathrm{crit}}^{(l=1)}\approx \allowbreak 1.56$ [see also Fig. \ref{fig6}(a)
below].

Differently from the GS, vortex modes are only partially stable, as
demonstrated by values of the eigenvalues for small perturbations, produced
by the numerical solution of BdG equations (\ref{chi_BdG}), and by direct
simulations of Eq. (\ref{varphi}) alike. We have performed a systematic
stability analysis for the vortex modes with $l=1$, in the case of $\sigma
=0 $ [no MF cubic term in Eq. (\ref{psi2d})]. First, it was found that, up
to accuracy of the numerically accumulated data, all eigenvalues have zero
real parts at $U_{0}>14/9$, and, in exact agreement with Eq. (\ref{14/9}),
pairs of unstable eigenvalues appear at $U_{0}<14/9$. Also in agreement with
the above derivation, the respective eigenmodes of small perturbations
correspond to $m=\pm 1$ in Eq. (\ref{U0<}), i.e., they are dipole modes,
which initiate spontaneous drift of the vortex' pivot off the central point.
An illustration of the resulting instability development is provided by Fig. %
\ref{fig6}(a), which shows that the drift instability triggers motion of the
center of mass (CM) of the vortex mode along the spiral trajectory, the CM's
location being defined as%
\begin{equation}
\left\{ x_{\mathrm{CM}},y_{\mathrm{CM}}\right\} =\frac{1}{N}\int \int
\left\{ x_{_{\mathrm{CM}}},y_{_{\mathrm{CM}}}\right\} \left\vert \psi \left(
x,y\right) \right\vert ^{2}dxdy,  \label{cm}
\end{equation}%
where $N$ is the total norm defined by Eq. (\ref{Npsi}). Eventually, the
pivot will be ousted to periphery, thus effectively converting the unstable
vortex mode into a stable GS with zero vorticity. In the case of $U_{0}>14/9$%
, the simulations demonstrate that a perturbed vortex with $l=1$ is a stable
mode which stays at the central position (not shown here in detail).

In the course of the simulations, a large part of the initial norm is
consumed by the absorber (emulating losses due to outward emission of
small-amplitude matter waves, in the indefinitely extended system). In
particular, the evolution of the unstable vortex displayed in Fig. \ref{fig6}%
(a) leads to its transformation into a residual GS with norm $N=1.88$ ($%
\approx 41\%$ of the initial value) and chemical potential $\mu \approx -95$%
, which is completely different from the initial value, $\mu =-1$. Taking
into regard that the transformation $l=1\rightarrow l=0$ implies the
replacement of $U_{l=1}=U_{0}-l^{2}=0.53$ by $U_{0}=1.53$, see Eq. (\ref{Ul}%
), it is worthy to note that the TF approximation, given by Eqs. (\ref{NTF})
with $\mu =-95$, yields a close value of the norm, $N_{\mathrm{TF}}^{(\sigma
=0)}(\mu )\approx 1.65$, even if $U_{0}=1.53$ is not a large value.

The spontaneous transformation of the vortex mode into the GS implies decay
of the mode's angular momentum, $L_{z}$, defined by Eq. (\ref{Lz}). In the
extended system, the momentum would be lost with emitted matter waves, while
in the present setting it is gradually eliminated by the edge absorber. The
dependence of $L_{z}$ on time, corresponding to the evolution of the
unstable vortex in Fig. \ref{fig6}(a), is shown in Fig. \ref{fig6}(b).

The spiral motion displayed in Fig. \ref{fig6}(a) represents only an initial
stage of the evolution [in the notation adopted in Fig. (\ref{psi2d}), the
time interval displayed in this figure is $t=7$]. At large times, when the
vortex's pivot will be lost in the periphery, the CM will eventually return
to the central position. In a fully conservative system, the CM would rather
orbit the center, cf. Eq. (\ref{omega}), but in the present setting the
effective dissipation induced by the absorber makes the return to the center
possible.

It is relevant to mention too that, as it follows from Eq. (\ref{U0<}), the
vortex mode with $l=1$ may become unstable against the perturbation with $%
m=2 $, i.e., against spontaneous splitting in two fragments, at still
smaller values of the pull strength, \textit{viz}., $U_{0}<5/9$. In this
work, we did not aim to detect this, apparently weaker, instability, in the
simulations. Lastly, for $l=2$ Eq. (\ref{lstability}) predicts a much
drift-instability region, $U_{0}<77/9$. This instability of the double
vortex could be easily detected in the simulations (not shown here in
detail).

\section{Conclusion}

While it was recently demonstrated that the quantum collapse, caused by the
potential of attraction to the center $\sim -r^{-2}$, can be suppressed by
the cubic repulsive MF (mean-field) nonlinearity in 3D bosonic gases, making
it possible to restore the otherwise missing GS (ground state), the cubic
self-repulsion is not sufficiently strong to stabilize the gas in the
effectively 2D setting. We have demonstrated that the effective quartic
repulsion, induced by the LHY\ (Lee-Huang-Yang) effect, i.e., the correction
to the MF theory produced by quantum fluctuations, provides the minimum
strength of nonlinearity sufficient for the stabilization of the 2D gas
under the action of the same attractive central potential. As a result, the
stable GS is created, with a singular but integrable density pattern. The
results are obtained in the numerical form, as well as by means of
analytical methods, based on the use of asymptotic expansions of the wave
functions at $r\rightarrow 0$ and $r\rightarrow \infty $, and TF
(Thomas-Fermi) approximation. A counterintuitive finding, an explanation to
which is given, is that the stable GS exists even in the case when the
central potential is repulsive, provided that its strength does not exceed a
critical value. In addition to the completely stable GS, partly stable
singular vortex states are constructed too, and their stability boundary is
found in an exact form. The pivot of an unstable vortex spontaneously drifts
away from the center along a spiral trajectory, the vortex being eventually
replaced by a stable GS. An estimate of the underlying physical parameters,
which correspond to the realization of the model in the form of the gas of
particles carrying permanent electric dipole moments, pulled to the central
charge, demonstrates that the radial size of the restored GS may be $R\sim 10
$ $\mathrm{\mu }$m, with the expected number of particles $\symbol{126}10^{5}
$.

As an extension of the present analysis, it may be interesting, in
particular, to analyze a setting with two mutually symmetric attraction
centers, a challenging possibility being prediction of spontaneous breaking
of the symmetry in the GS pinned to the pair of the centers.

\section*{Acknowledgment}

This work was supported, in part, by the Israel Science Foundation through
grant No. 1286/17.

\end{document}